\documentclass[prl,twocolumn,showpacs,superscriptaddress,amsmath,amssymb]{revtex4}

\usepackage[utf8]{inputenc}

\usepackage{graphicx}
\usepackage[colorlinks=true,citecolor=blue]{hyperref}
\usepackage{verbatim}

\usepackage{units}
\usepackage{color}
\usepackage{ulem}

\newcommand{\kBT}{k_\text{B}T}
\newcommand{\kB}{k_\text{B}}
\newcommand{\bea}{\begin{eqnarray}}
\newcommand{\eea}{\end{eqnarray}}
\newcommand{\be}{\begin{equation}}
\newcommand{\ee}{\end{equation}}

\begin{document}

\title{
\vspace*{-1.25cm}
\textnormal{{\small \flushright PHYSICAL REVIEW LETTERS {\bf 114}, 146801 (2015)}}\\
\vspace*{-0.2cm}
\rule[0.1cm]{18cm}{0.02cm}\\
\vspace*{0.285cm}
Chiral thermoelectrics with quantum Hall edge states}
\author{Rafael S\'anchez}
\affiliation{Instituto de Ciencia de Materiales de Madrid, CSIC, Cantoblanco, 28049, Madrid, Spain}
\author{Bj\"orn Sothmann}
\affiliation{D\'epartement de Physique Th\'eorique, Universit\'e de Gen\`eve, CH-1211 Gen\`eve 4, Switzerland}
\author{Andrew N. Jordan}
\affiliation{Department of Physics and Astronomy, University of Rochester, Rochester, New York 14627, USA}
\affiliation{Institute for Quantum Studies, Chapman University, Orange, California 92866, USA}

\begin{abstract}
The thermoelectric properties of a three-terminal quantum Hall conductor are investigated. We identify a contribution to the thermoelectric response that relies on the chirality of the carrier motion rather than on spatial asymmetries. The Onsager matrix becomes maximally asymmetric with configurations where either the Seebeck or the Peltier coefficients are zero while the other one remains finite. Reversing the magnetic field direction exchanges these effects, which originate from the chiral nature of the quantum Hall edge states. The possibility to generate spin-polarized currents in quantum spin Hall samples is discussed.
\end{abstract}
\maketitle 

The integer quantum Hall effect occurs in two-dimensional conductors subject to a strong perpendicular magnetic field. It manifests itself as quantized plateaus of the Hall conductance~\cite{v._klitzing_new_1980}. Theoretically, this can be understood in terms of dissipationless transport along chiral edge states within Landauer-B\"uttiker theory~\cite{halperin_quantized_1982,buttiker_absence_1988}.  Thermal transport along these edge channels was recently investigated to trace the path of energy relaxation along the edge~\cite{granger_observation_2009,nam_thermoelectric_2013} or probe the presence of neutral modes~\cite{altimiras_non-equilibrium_2010}, carrying energy but not charge, which are not accessible in electrical current measurements. However, until now there is surprisingly little known about thermoelectric properties in the integer quantum Hall regime.


Thermoelectric properties of multiterminal conductors have recently received a lot of attention~\cite{sothmann_thermoelectric_2014} with a particular emphasis on nonlinear effects~\cite{sanchez_scattering_2013,whitney_thermodynamic_2013,meair_scattering_2013}. In these systems, a hot terminal injects heat but no charge into the conductor, driving a directed charge current between two other cold terminals. Charge and heat flows are thus separated, allowing the electrical circuit driving a load to be at a single temperature.  The heat source can be of fermionic~\cite{sanchez_optimal_2011,jordan_powerful_2013} or bosonic nature~\cite{entin-wohlman_three-terminal_2010,sothmann_magnon-driven_2012,bergenfeldt_hybrid_2014}. A crossed thermopower appears by heat rectification which, in the absence of a magnetic field, depends on the breaking of both left-right and particle-hole symmetries.


In the presence of a magnetic field, the off-diagonal elements of the thermoelectric linear-response Onsager matrix are unrelated to each other due to the broken time-reversal symmetry~\cite{onsager_reciprocal_1931,buttiker_symmetry_1988,butcher_thermal_1990}. This property relaxes the (broken) symmetry requirements discussed above. The absence of backscattering along edge channels, a property of the quantum Hall effect \cite{buttiker_absence_1988}, maximizes this broken reciprocity, as we discuss here. In particular, we show that chiral propagation
introduces a unique transport feature such that while the Seebeck coefficient is finite, the Peltier one can be zero, or vice versa. This effect is a consequence of the edge states, and can be used to generate spin-polarized currents in topological insulators~\cite{kane_z2_2005,bernevig_quantum_2006,konig_quantum_2007}.

The ramifications of the asymmetry in the Onsager matrix on the thermoelectric performance of mesoscopic heat engines is an active research topic~\cite{benenti_thermodynamic_2011,sanchez_thermoelectric_2011,saito_thermopower_2011,brandner_strong_2013,brandner_multi-terminal_2013,stark_classical_2014,sothmann_quantum_2014}. It has been shown that broken time-reversal symmetry in principle allows for Carnot efficiency $\eta_C$ at {\it maximal output power}~\cite{benenti_thermodynamic_2011,saito_thermopower_2011}. However, additional constraints from current conservation in a multiterminal setup restrict the efficiency at maximum power to values smaller than $\eta_C$ where the precise bound depends on the number of terminals~\cite{brandner_strong_2013,brandner_multi-terminal_2013}. As first examples, classical and quantum Nernst engines have been shown to saturate the efficiency bounds under their particular boundary conditions~\cite{stark_classical_2014,sothmann_quantum_2014}. Here, we also demonstrate that more flexible and experimentally feasible boundary conditions can actually increase these limits even further.


\begin{figure}[b]
\begin{center}
\includegraphics[width=\linewidth,clip] {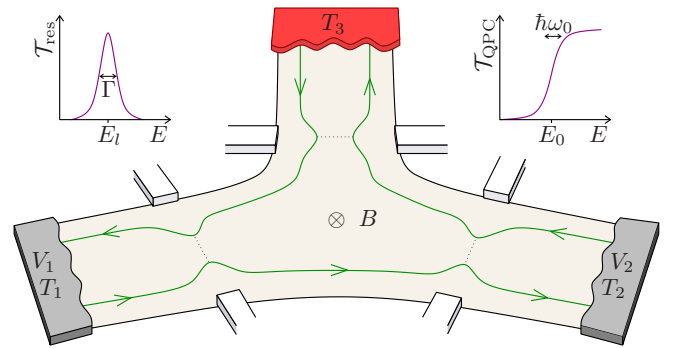}
\vspace{-0.8cm}
\end{center}
\caption{\label{scheme} Three-terminal quantum Hall bar. A finite current is generated along the edge state between cold terminals 1 and 2 by conversion of heat injected from the hot probe terminal 3, originating from a temperature bias $\Delta T_3$. Details of the energy-dependent scattering at the constrictions influence the thermoelectric response dramatically, revealing the chiral nature of electronic propagation in the sample.
}
\vspace{-0.15cm}
\end{figure}

We focus on the simplest configuration with chiral thermoelectric properties consisting of a three-terminal Hall bar at filling factor $\nu=1$, as depicted in Fig.~\ref{scheme}. Two terminals are at the same (cold) temperature $T_1=T_2=T$ and support a charge current, $I^e$. The third terminal is hot, having a temperature bias $\Delta T_3=T_3-T$, and is considered to be a voltage probe~\cite{buttiker_coherent_1988,bergfield_thermoelectric_2014} such that it injects no net charge current into the conductor.
Constrictions in each arm give rise to energy-dependent scattering of edge channels ---a prerequisite for thermoelectric response. The transmission coefficients at each constriction (e.g. a resonance or a saddle point) determine the details of the thermoelectric performance.

Transport through the system is described within the Landauer-B\"uttiker formalism~\cite{buttiker_absence_1988}. Charge and heat currents ${\mathbf I}_i=(I_i^e,I_i^h)$ are given by the transmission probabilities ${\cal T}_{i\leftarrow j}(E)$ for electrons injected in terminal $j$ to be absorbed by terminal $i$. Generally, the currents are written in linear response as
\be
\label{Ii}
{\mathbf I}_i=
\frac{1}{h}\sum_j\int dE[\delta_{ij}-{\cal T}_{i\leftarrow j}(E)]\xi(E)
\left(
	\begin{array}{cc}
		e & eE \\
		E & E^2
	\end{array}
\right)
{\mathbf F}_j,
\ee
in terms of the electric and thermal affinities ${\mathbf F}_j=(F_j^V,F_j^T)$, with $F_i^V=eV_i/(\kBT)$ and $F_i^T=k_\text{B}\Delta T_i/(\kBT)^2$,
where $V_i$ is the voltage applied to terminal $i$. We have defined $\xi(E)=-(\kBT/2)df(E)/dE$, with the Fermi function $f(E)$. We choose the equilibrium Fermi energy as the zero of energy.

Assuming the boundary conditions $F_1^T=F_2^T=0$, we solve Eqs.~\eqref{Ii} to get the voltage developed at the probe satisfying $I_3^e=0$. From charge conservation we thus have $I^e=I_1^e=-I_2^e$. We can then write the relevant Onsager coefficients ${\mathbf {\cal L}}$ for the generated charge current and the absorbed heat, 
\be
\left(
	\begin{array}{c}
		I^e \\
		I_3^h
	\end{array}
\right)
=
\left(
	\begin{array}{cc}
		{\cal L}_{eV} & {\cal L}_{eT} \\
		{\cal L}_{hV} & {\cal L}_{hT}
	\end{array}
\right)
\left(
	\begin{array}{c}
		F_1^V{-}F_2^V \\
		F_3^T
	\end{array}
\right).
\ee
The diagonal terms of the Onsager matrix correspond to the charge and heat conductances, while the off-diagonal ones are related to the crossed Seebeck (${\cal L}_{eT}$) and Peltier (${\cal L}_{hV}$) coefficients. Their ratio $x={\cal L}_{eT}/(e{\cal L}_{hV})$ can deviate from one in the presence of a magnetic field. 

Let us consider the configuration sketched in Fig.~\ref{scheme}, where the cold terminals 1 and 2 are constricted (${\cal T}_{1}(E),  {\cal T}_{2}(E) \leq 1$), but the hot terminal 3 is open (${\cal T}_{3}(E) =1$).  Hence, electrons injected from the hot terminal at energy $E$ are reabsorbed only after being reflected at the two junctions with probability ${\cal T}_{3\leftarrow 3}=(1-{\cal T}_1(E))(1-{\cal T}_2(E))$. 
We obtain for the crossed thermoelectric coefficients 
\begin{align}
\label{Let}
{\cal L}_{eT}&=k_\text{B}T^2G(S_2-S_1)+e{\cal X}_1,\\
\label{Lhv}
{\cal L}_{hV}&=e^{-1}{\cal L}_{eT}-{\cal X}_1-{\cal X}_2,
\end{align}
where the conductance $G=\Lambda^{-1}/(G_1^{-1}{+}G_2^{-1})$ is modified from that of a sequencial transmission of two barriers~\cite{buttiker_coherent_1988} ---each with an individual conductance $G_l=e^2(\kBT h)^{-1}\int dE{\cal T}_l(E)\xi(E)$--- by the factor $\Lambda=1-J_1/(G_1{+}G_2)$. We define the integrals $J_n=e^2(h\kBT)^{-1}\int dE E^{n-1}{\cal T}_1(E){\cal T}_2(E)\xi(E)$ which are due to the coherent propagation between the two junctions. The first term in Eq.~\eqref{Let}, depending on the difference of the two terminal thermopower of each junction, $S_l=e/(hk_\text{B}T^2G_l)\int dEE{\cal T}_l(E)\xi(E)$, is expected for three-terminal heat rectifiers~\cite{sothmann_thermoelectric_2014,mazza_thermoelectric_2014}. Remarkably, the additional term,
\be
{\cal X}_l=\frac{\kBT}{e^2}\frac{GG_{l}}{G_1G_2}(eTS_lJ_1-J_2),
\label{X}
\ee
introduces a dependence on the direction of propagation, which is determined by the sign of the magnetic field: the index $l$ denotes the junction that is hit by the electrons injected from the hot terminal. This term relies on the coherent propagation between the two junctions, as implicitly included in the dependence on $J_n$. Note furthermore that ${\cal L}_{eT}(B)-{\cal L}_{eT}(-B)=e({\cal X}_1+{\cal X}_2)$.  Eqns.~(\ref{Let},\ref{Lhv},\ref{X}) are our main results, which we will now expound.

\begin{figure}[t]
\begin{center}
\includegraphics[width=\linewidth,clip] {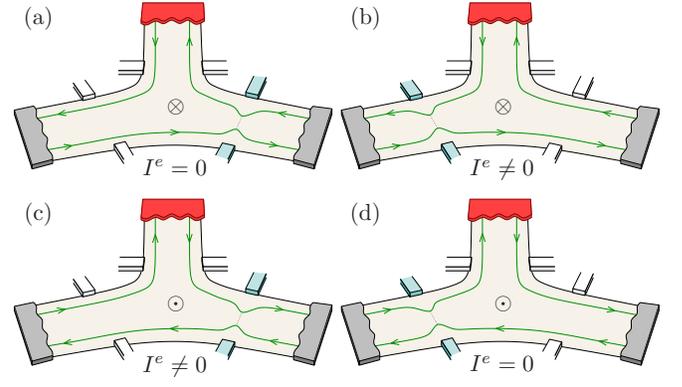}
\vspace{-0.6cm}
\end{center}
\caption{\label{gateB} Different configurations for which chirality manifests as a generated current being finite (b,c) or zero (a,d) depending on the position of the junction and the sign of the magnetic field.
}
\vspace{-0.2cm}
\end{figure}

We emphasize the effect of chirality in the thermoelectric response (entering through ${\cal X}_l$) by considering the configurations with one of the junctions totally transmitting. {\it (i)} If  ${\cal T}_1(E)=1$, cf. Fig.~\ref{gateB}(a), we have $S_1=0$. Then on one hand, the chiral term ${\cal X}_1=-e^{-1}k_\text{B}T^2GS_2$ cancels the first term in Eq.~\eqref{Let}, resulting in ${\cal L}_{eT}=0$. 
On the other hand, ${\cal L}_{hV}=e^{-1}k_\text{B}T^2G_2S_2$: with no temperature bias we recover the Peltier response of a two-terminal configuration. {\it (ii)} However, in the opposite case, ${\cal T}_2(E)=1$, cf. Fig.~\ref{gateB}(b), we have ${\cal X}_1=0$ and the response is proportional to the thermopower of the other barrier ${\cal L}_{eT}=-k_\text{B}T^2G_1S_1$, with ${\cal L}_{hV}=0$. 
The different Seebeck matrix elements can be understood as follows. In case {\it(i)}, the hot terminal creates electron-hole excitations in the stream of particles emitted from terminal 2. As these excitations are neutral, there is no net charge current in the system. However, in case {\it(ii)}, the particle-hole excitations created by the hot terminal get partitioned before entering terminal 1, thus giving rise to a net charge current.
Remarkably in both cases, while either the Seebeck or the Peltier coefficient is similar to that of a two-terminal system, the other one is exactly zero. The two configurations are interchanged when reversing the magnetic field ---Figs.~\ref{gateB}$($c) and (d)---, as follows from Onsager reciprocity relations. This effect of having either the presence or absence of a Seebeck (or Peltier) effect depending on the sign of the magnetic field is a consequence of systems with chiral propagation along edge states. It can therefore be used to probe their properties in a sample by tuning the sign of the magnetic field and the gate voltages that open/close the two junctions. 

This result also applies to quantum spin Hall topological insulators where a strong spin-orbit interaction plays the role of the magnetic field. In that case, helical edge states carry electrons with opposite spin along opposite directions~\cite{kane_z2_2005,bernevig_quantum_2006,konig_quantum_2007}. Thus, the discussion in the previous paragraph will apply in each case (with either of the junctions open) to a different spin component. In this case, the charge current is in general finite even if one of the junctions is open: If say spin up electrons move clockwise and spin down, counter-clockwise and junction 1 is open, only spin up electrons will contribute to the current.
Thus, our system can be operated as a {\it spin polarizer}~\cite{dolcetto_generating_2013,hwang_nonlinear_2014}, with the polarization of the generated current determined by which of the two junctions is open.

\begin{figure}[t]
\begin{center}
\includegraphics[width=0.9\linewidth,clip] {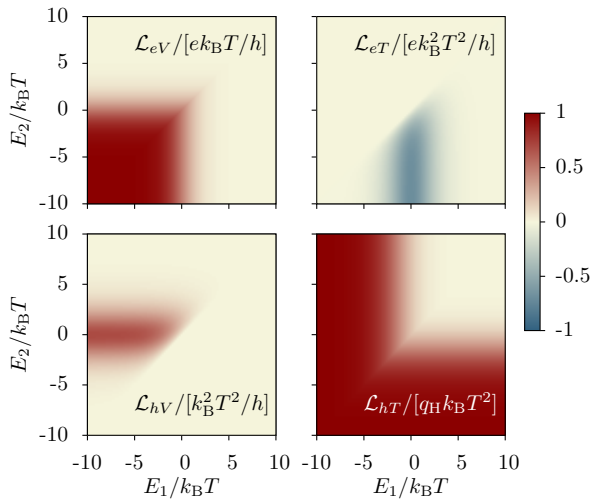}
\vspace{-0.4cm}
\end{center}
\caption{\label{onsager} Representation of the Onsager matrix ${\mathbf{\cal L}}$ for a system consisting of two QPCs with threshold energies $E_1$ and $E_2$ and $\hbar\omega_0=\kBT/10$ in front of the conducting terminals. 
}
\vspace{-0.2cm}
\end{figure}

We illustrate the dependence on the scattering details in Fig.~\ref{onsager} for the case where the two junctions are quantum point contacts (QPCs). They are described by a saddle-point potential with transmission probability~\cite{buttiker_quantized_1990} ${\cal T}_\text{QPC}(E)=[1+e^{-2\pi(E-E_l)/\hbar\omega_{0,l}}]^{-1}$. Here, $E_l$ and $\omega_{0,l}$ are the position and width of the step of junction $l$, respectively. Junctions with $-E_l\gg\hbar\omega_{0,l}$ are open. Thus, the conductance ${\cal L}_{eV}$ vanishes when at least one of the junctions is closed, with the injected thermal conductance ${\cal L}_{hV}$ requiring at least one of the junctions to be open in order to be finite. The Seebeck matrix element ${\cal L}_{eT}$ shows a dip when QPC 1 is tuned to threshold as expected~\cite{molenkamp_quantum_1990}. In agreement with the discussion above, it is finite only when terminal 1 is partially open (noisy) and terminal 2 is not closed. For typical experimental temperatures around 100 mK the Seebeck effect will generate a measurable current of around 2 nA for a gradient of 10 mK, which is readily detectably with current technology. Similarly, the Peltier matrix element ${\cal L}_{hV}$ exhibits a peak when QPC 2 is at threshold. Thus, chiral edge states allow the asymmetry of the Onsager matrix, $x$, to be tuned through the whole range from plus to minus infinity by simply changing the gate voltages and the magnetic field.

\begin{figure}[t]
\begin{center}
\includegraphics[width=\linewidth,clip] {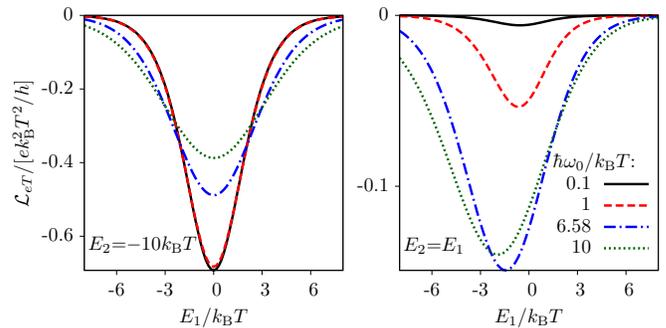}
\vspace{-0.8cm}
\end{center}
\caption{\label{LeT} Thermoelectric response, ${\cal L}_{eT}$, for the same configuration of Fig.~\ref{onsager}. If the right QPC is open, $E_2=-10\kBT$, we have ${\cal X}_1=0$. Then,  increasing the width $\omega_0$ smears the energy dependence of the left junction and the maximum of the current is monotonically decreased. A different behavior shows up for a symmetric configuration with $E_1=E_2$ where only the term $e{\cal X}_1$ contributes: ${\cal L}_{eT}\to0$ for $\omega_0\to0$, with a maximum at $\hbar\omega_0\approx6.6\kBT$. 
}
\vspace{-0.2cm}
\end{figure}

Another important consequence of Eq.~\eqref{Let} is that it relaxes the symmetry-breaking conditions to have a finite thermoelectric response. In the absence of a magnetic field, a Peltier or Seebeck effect requires both broken left-right and particle-hole symmetries. As we discussed above, the condition on broken electron-hole symmetry (energy dependent transmission) is put on only one of the junctions. The other one can be totally open. Additionally, even in the case of having a left-right symmetric system, we can obtain a finite thermopower. In this case, with ${\cal T}_1(E)={\cal T}_2(E)$, we have ${\cal L}_{eT}=e{\cal X}_1$, i.e. it is finite only due to the contribution of the magnetic field. Remarkably at this symmetric configuration, the Onsager reciprocity ${\cal L}_{eT}=e{\cal L}_{hV}$ is satisfied without needing to reverse the magnetic field.  This must be the case because a reversal of the magnetic field is physically equivalent to reflecting the physical system about its symmetry axis.

The chiral term ${\cal X}_1$ introduces a non-trivial dependence of the Seebeck matrix element on the sharpness of the transmission step, as shown in Fig.~\ref{LeT}. If the right junction is open, i.e. ${\cal X}_1=0$, the response decreases with the smearing of the energy dependence of the ${\cal T}_1(E)$ with $\omega_0$. On the other hand, for a symmetric configuration with $E_1=E_2$, such that only ${\cal X}_1$ contributes, the behavior is different: for $\omega_0\rightarrow0$, the transmission is given by a Heaviside step function and ${\cal L}_{eT}\rightarrow0$. A maximal value is found for widths not much larger than the thermal smearing $\kBT$.  

We gain further insight by considering weakly energy-dependent junctions. In that regime, the two-terminal thermopower is given by
a Cutler-Mott like formula as a logarithmic derivative of the conductance~\cite{cutler_observation_1969,lunde_Mott_2005}. In our case, the crossed thermopower $S_{\cal X}={\cal L}_{eT}/(k_\text{B}TG)$ leads in a Sommerfeld expansion to
\be
\label{sx}
S_{\cal X}=\frac{q_Hh}{e}\left[-\partial_E\ln{{\cal T}_1}+(1-{\cal T}_1)\partial_E\ln{{\cal T}_2}\right]_{E=E_\text{F}},
\ee
where we have introduced the quantum of heat conductance~\cite{pendry_quantum_1983} $q_H=\pi^2k_\text{B}^2T/(3h)$. 
This formula separates the contribution of the scattering of electrons injected from the hot terminal at the two junctions. The different role of the two junctions is a consequence of the chiral propagation. The first term is the Cutler-Mott like (two-terminal) thermopower of the left junction. Remarkably, the contribution of the second junction is weighted by the reflection probability at the first one. This is intuitive because only those hot electrons that are reflected at the left junction are scattered by the second one. The left junction modifies the population of the edge channel incident to the right one. The relative sign is due to the central position of the heat source, so the two junctions have opposite contributions to the crossed thermopower. If the magnetic field is reversed, the roles of terminal 1 and 2 exchange.

\begin{figure}[t]
\begin{center}
\includegraphics[width=0.9\linewidth,clip] {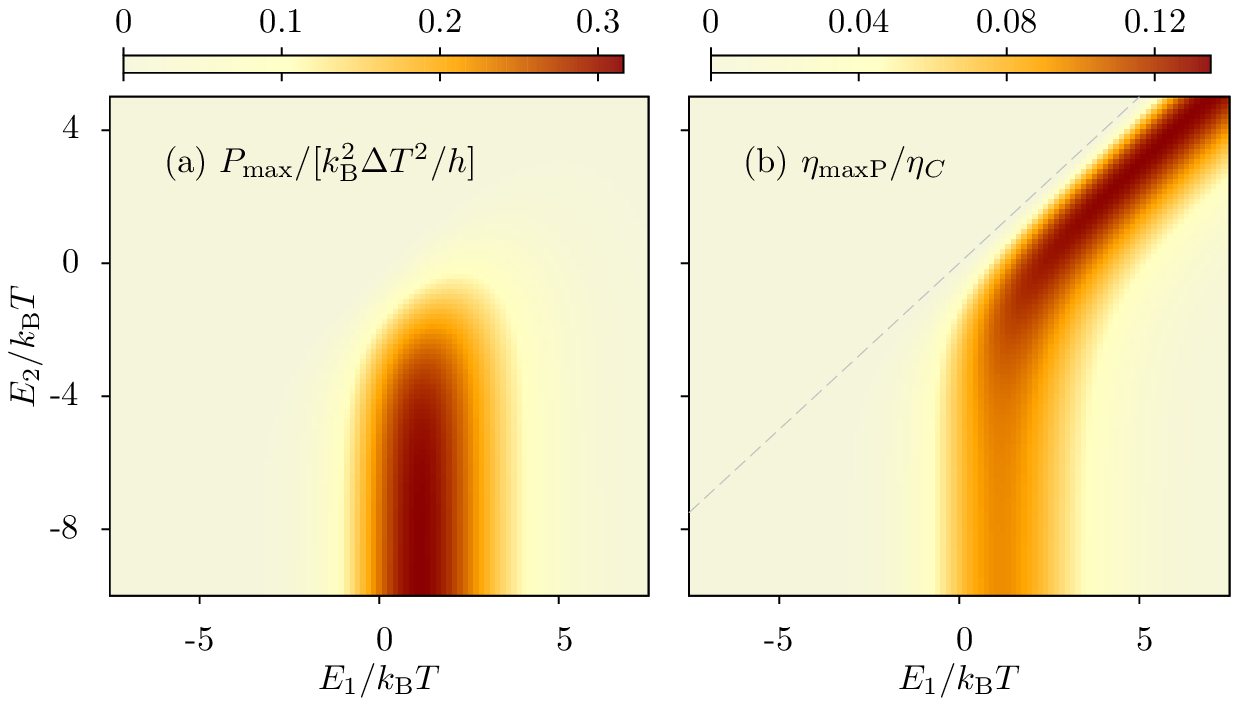}
\includegraphics[width=0.9\linewidth,clip] {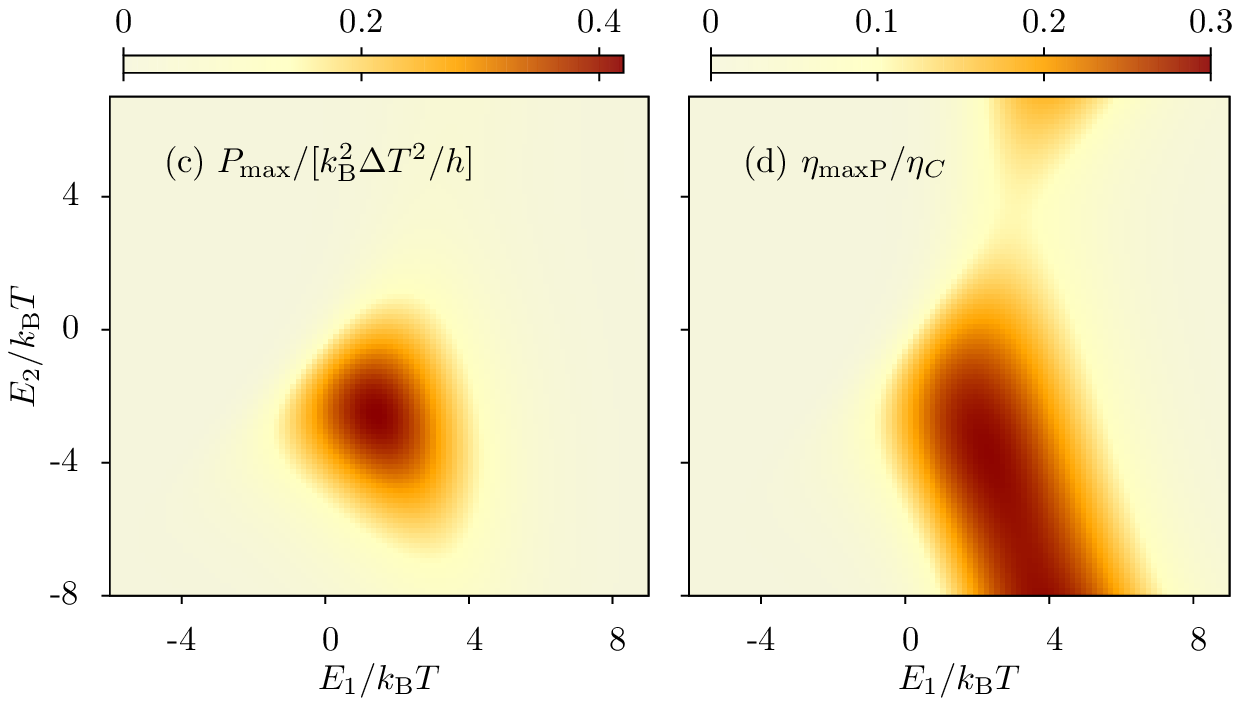}
\vspace{-0.4cm}
\end{center}
\caption{\label{emP} Maximum power and efficiency at maximum power, for the two different configurations with (a),(b) a QPC in junctions 1 and 2 and (c),(d) a QPC in junction 1 and a resonance in junction 2. The QPCs have sharp transmission steps with width $\hbar\omega_0=0.1\kBT$ and threshold energy $E_l$. The resonance width $\Gamma=2.1\kBT$ in (c),(d) is optimized to give the maximal $P_\text{max}$.
}
\vspace{-0.3cm}
\end{figure}

We finally turn to a discussion of the thermoelectric performance of our quantum Hall heat engine. Two important characterizations are the output power $P=I^e(V_2-V_1)$ maximized with respect to the applied bias voltage for a given temperature bias, $P_\text{max}=\kBT{\cal L}_{eT}^2(F_3^T)^2/(4e{\cal L}_{eV})$, and the efficiency at maximum power, $\eta_\text{maxP}=\eta_Cxy/(4+2y)$, with $y={\cal L}_{eT}{\cal L}_{hV}/\det{\cal L}$. 
In Figs.~\ref{emP}(a) and (b) we show the maximum power and $\eta_\text{maxP}$ for such a configuration with two QPCs as the centers of the energy steps are varied around the Fermi energy. The maximal $\eta_\text{maxP}$ is found in the region where the two QPCs are above the Fermi level, with $E_1-E_2\approx2\kBT$. There only the very few hottest electrons from terminal 3 contribute to transport. All the rest are reflected by both junctions. Thus the power extraction is negligible. A good compromise for a high power and efficiency is found in the region where $E_1$ and $-E_2$ are of the order of $\kBT$, where we find $P_\text{max}\approx 0.3(\kB\Delta T)^2/h$ and $\eta_\text{maxP}\approx0.1\eta_C$ comparable to resonant tunneling heat engines~\cite{jordan_powerful_2013,sothmann_powerful_2013}.

An even larger output power can be found when resonant tunneling occurs in one of the junctions, e.g., in the presence of an antidot~\cite{sim_electron_2008}. In that case, the transmission is described by~\cite{buttiker_coherent_1988} ${\cal T}_\text{res}(E)=\Gamma_l^2[4(E-E_l)^2+\Gamma_l^2]^{-1}$. In Figs.~\ref{emP}$($c) and (d) we show the thermoelectric performance at a resonance width $\Gamma\approx2.1\kBT$ optimized for maximal $P_\text{max}$. Remarkably, the maxima of $P_\text{max}\approx 0.4(\kB\Delta T)^2/h$ and $\eta_\text{maxP}\approx 0.3\eta_C$ coincide for $E_1$ and $E_2$ close to the Fermi level. Importantly, we can search for parameters that optimize the efficiency at maximum power.  For $\Gamma\to 0$, an upper limit $\eta_\text{maxP}=\eta_C/2$ is reached. Thus, a three-terminal quantum Hall engine can outperform a quantum Nernst engine~\cite{sothmann_quantum_2014} by a factor of two in terms of efficiency at maximum power.

In conclusion, we have demonstrated that three-terminal thermoelectrics is sensitive to the chiral properties of electronic propagation in quantum Hall systems. An additional term that accounts for the direction of the electron motion rather than geometric symmetries appears in the usual Seebeck coefficient. In particular, this term is responsible for the possibility to switch off either the Seebeck or the Peltier response by tuning the scattering details. The mechanism can be also used to generate spin polarized currents in topological insulators. Furthermore, the system also works as a powerful and efficient energy harvester.

We acknowledge support from the Spanish MICINN Juan de la Cierva program and MAT2011-24331, COST Action MP1209, and the Swiss National Science Foundation via the NCCR QSIT.

%

\end{document}